\documentclass[preprint,authoryear,3p,12pt,times]{elsarticle}  % so abschicken

\usepackage[T2A,T1]{CJKutf8}

\usepackage[ukrainian,russian,british]{babel}

\usepackage{epsfig}
\usepackage{amssymb}

\newcommand{\qq}[1]{\lq{#1}\rq}

\newcommand{\chn}[1]{\bgroup\begin{CJK*}{UTF8}{}\CJKtilde\CJKfamily{gbsn}{#1}\end{CJK*}\egroup}

\newcommand{\absnum}{\ensuremath{\mathnormal{N}}}

\newcommand{\area}{\ensuremath{\mathnormal{F}}}
\newcommand{\chem}[1]{\ensuremath{\mathrm{#1}}}
\newcommand{\chempot}{\ensuremath{\mathnormal{\mu}}}
\newcommand{\CNT}[1]{\ensuremath{{#1}_\mathrm{PE}}}
\newcommand{\CNTnop}[1]{\ensuremath{{#1}_\mathrm{SC}}}
\newcommand{\committor}{\ensuremath{\mathcal{Q}}}
\newcommand{\component}{\ensuremath{\mathnormal{i}}}
\newcommand{\critical}[1]{\ensuremath{{#1}^\mathnormal{\star}}}

\newcommand{\Deltamu}{\ensuremath{\mathnormal{\Delta\chempot}}}
\newcommand{\Deltamueff}{\ensuremath{\Deltamu_\mathrm{e}}}
\newcommand{\density}{\ensuremath{\mathnormal{\rho}}}
\newcommand{\densitymon}{\ensuremath{\tilde{\density}}}
\newcommand{\DGcritCNT}{\ensuremath{\Delta\gibbs^\mathnormal{\star}_\mathrm{PE}}}
\newcommand{\differential}{\ensuremath{\mathnormal{d}}}

\newcommand{\echarge}{\ensuremath{\mathnormal{e}}}
\newcommand{\elongation}{\ensuremath{\mathnormal{\ell}}}
\newcommand{\enthalpy}{\ensuremath{\mathnormal{h}}}

\newcommand{\federb}{\ensuremath{\mathnormal{b}}}
\newcommand{\federq}{\ensuremath{\mathnormal{q}}}
\newcommand{\CNTfederbb}{\ensuremath{\mathnormal{b^2_\mathrm{PE}}}}
\newcommand{\CNTfederqq}{\ensuremath{\mathnormal{q^2_\mathrm{PE}}}}
\newcommand{\fluidA}{\ensuremath{\mathrm{A}}}
\newcommand{\fluidB}{\ensuremath{\mathrm{B}}}
\newcommand{\formationrate}{\ensuremath{\mathnormal{I}}}
\newcommand{\gaseffect}{\ensuremath{\mathcal{G}}}
\newcommand{\gibbs}{\ensuremath{\mathnormal{G}}}
\newcommand{\heatcapacity}{\ensuremath{\mathnormal{c}}}
\newcommand{\invsupfraction}{\ensuremath{\mathnormal{\mathnormal{Y}_{0}^{-1}}}}
\newcommand{\isochoric}[1]{\ensuremath{\heatcapacity_\mathnormal{v, {#1}}}}
\newcommand{\kboltz}{\ensuremath{\mathnormal{k}}}
\newcommand{\landau}[1]{\ensuremath{\mathcal{O}\left({#1}\right)}}
\newcommand{\liq}[1]{{#1}'}
\newcommand{\liqfraction}{\ensuremath{\mathnormal{x}}}
\newcommand{\LJenergy}{\ensuremath{\mathnormal{\varepsilon}}}
\newcommand{\LJlength}{\ensuremath{\mathnormal{\sigma}}}
\newcommand{\mass}{\ensuremath{\mathnormal{m}}}
\newcommand{\molarfraction}[1]{\ensuremath{\mathnormal{Y}_\mathnormal{#1}}}

\newcommand{\mult}{\ensuremath{\times}}
\newcommand{\NcritCNT}{\ensuremath{\nuclsize^\mathnormal{\star}_\mathrm{PE}}}
\newcommand{\nuclrate}{\ensuremath{\mathnormal{j}}}
\newcommand{\nuclrateiso}{\ensuremath{\nuclrate_\temperature}}
\newcommand{\CNTnopnuclrateiso}{\ensuremath{\nuclrate_{\temperature, \mathrm{SC}}}}
\newcommand{\CNTnuclrateiso}{\ensuremath{\nuclrate_{\temperature, \mathrm{PE}}}}
\newcommand{\nuclsize}{\ensuremath{\mathcal{N}}}
\newcommand{\numcomponents}{\ensuremath{\mathnormal{K}}}
\newcommand{\partialpressure}{\ensuremath{\mathcal{P}}}
\newcommand{\planartension}{\ensuremath{\surfacetension_\mathnormal{0}}}
\newcommand{\pressure}{\ensuremath{\mathnormal{p}}}
\newcommand{\quadrupole}{\ensuremath{\mathnormal{Q}}}
\newcommand{\radius}{\ensuremath{\mathcal{R}_\surfacetension}}
\newcommand{\sat}[1]{\ensuremath{{#1}_\mathrm{s}}}
\newcommand{\supersatrho}{\ensuremath{\mathnormal{S}}}
\newcommand{\sps}{\ensuremath{\supersatrho^\mathnormal{\nabla}}}
\newcommand{\supfraction}{\ensuremath{\mathnormal{\molarfraction{0}}}}
\newcommand{\supmass}{\ensuremath{\mass_\mathnormal{0}}}
\newcommand{\surfacetension}{\ensuremath{\mathnormal{\gamma}}}
\newcommand{\Tcrit}{\ensuremath{\temperature_\mathrm{c}}}
\newcommand{\temperature}{\ensuremath{\mathnormal{T}}}

\newcommand{\threshold}{\ensuremath{\mathnormal{M}}}
\newcommand{\tolmanlength}{\ensuremath{\mathnormal{\delta}}}

\newcommand{\trate}{\ensuremath{\mathnormal{\beta}}}
\newcommand{\unlike}{\ensuremath{\mathnormal{\xi}}}
\newcommand{\vap}[1]{{#1}''}
\newcommand{\vapfraction}{\ensuremath{\mathnormal{y}}}

\newcommand{\volume}{\ensuremath{\mathnormal{V}}}
\newcommand{\Weffect}{\ensuremath{\mathcal{W}}}
\newcommand{\zeldovich}{\ensuremath{\mathnormal{Z}}}

\newcommand{\txtCNTnop}{\textnormal{SC}}  % simplified CNT
\newcommand{\txtCNT}{\textnormal{PE}}  % CNT (with the pressure effect)

\newcommand{\oou}{ou}

\newcommand{\vapor}{vap\oou{}r}
\newcommand{\vapour}{\vapor}
\newcommand{\Vapor}{Vap\oou{}r}
\newcommand{\Vapour}{\Vapor}
\newcommand{\vapors}{vap\oou{}rs}
\newcommand{\vapours}{\vapors}

\newcommand{\nucleus}{droplet}
\newcommand{\Nucleus}{Droplet}
\newcommand{\nuclei}{droplets}

\newcommand{\droplet}{\nucleus}
\newcommand{\Droplet}{\Nucleus}
\newcommand{\droplets}{\nuclei}

\biboptions{comma}

\journal{Atmospheric Research}

\begin{document}
\begin{frontmatter}

\title{The air pressure effect on the homogeneous nucleation of carbon dioxide by molecular simulation}
\author[Paderborn]{M.\ Horsch}
\author[Paderborn]{Z.\ Lin\fnref{lin}}
\fntext[lin]{\chn{蔺增勇} (Lin Zengyong)}
\author[Paderborn]{T.\ Windmann}
\author[Kaiserslautern]{H.\ Hasse}
\author[Paderborn]{J.\ Vrabec\corref{cor1}}
\cortext[cor1]{Author to whom correspondence should be addressed: Jadran Vrabec, Universit\"at Paderborn, Institut f\"ur Verfahrens\-technik, Lehrstuhl f\"ur Thermodynamik und Energietechnik, +49 5251 60 2421 (phone), +49 5251 60 3522 (fax).}
\ead{jadran.vrabec@upb.de}
\ead[url]{http://thet.upb.de/}

\address[Paderborn]{Universit\"at Paderborn, Lehrstuhl f\"ur Thermodynamik und Energietechnik, Warburger Str.\ 100, 33098 Paderborn, Germany}
\address[Kaiserslautern]{Technische Universit\"at Kaiserslautern, Lehrstuhl f\"ur Thermodynamik, Erwin-Schr\"odinger-Str.\ 44, 67663 Kaiserslautern, Germany}

\begin{abstract}
\Vapour-liquid equilibria (VLE) and the influence of an inert carrier gas on homogeneous
\vapour{} to liquid nucleation are investigated by molecular simulation for quaternary mixtures
of carbon dioxide, nitrogen, oxygen, and argon.
Canonical ensemble molecular dynamics simulation using the Yasuoka-Matsumoto method
is applied to nucleation in supersaturated \vapours{} that contain more
carbon dioxide than in the saturated state at the dew line.
Established molecular models are employed that are known to accurately reproduce the VLE
of the pure fluids as well as their binary and ternary mixtures.
On the basis of these models, also
the quaternary VLE properties of the bulk fluid
are determined with the Grand Equilibrium method.

Simulation results for the carrier gas influence on the nucleation rate are compared
with the classical nucleation theory (CNT) considering the
\qq{pressure effect} [Phys.\ Rev.\ Lett.\ 101: 125703 (2008)].
It is found that the presence of air as
a carrier gas decreases the nucleation rate only slightly
and, in particular, to a significantly lower extent than predicted by CNT.
The nucleation rate of carbon dioxide is generally underestimated by CNT, leading to
a deviation between one and two orders of magnitude for pure carbon dioxide in
the vicinity of the spinodal line and up to three orders of magnitude in presence
of air as a carrier gas. Furthermore, CNT predicts a temperature dependence of the
nucleation rate in the spinodal limit, which cannot be confirmed by molecular simulation.
\end{abstract}

\end{frontmatter}

%% main text
\section{Introduction}
\label{sec:int}

If significant reductions in greenhouse gas emissions are to be achieved,
it is essential to minimize the energy required for separating carbon dioxide
from air as a prerequisite for rendering purification, recovery and
sequestration processes economically and ecologically more
efficient \citep{Kaule02}. This requires an accurate understanding of the phase
behavi\oou{}r of carbon dioxide (\chem{CO_2}) in the earth's atmosphere under diverse
conditions, corresponding to the respective technical applications.

Condensation processes, initiated by nucleation, as well as \vapour{}-liquid
equilibria (VLE) are particularly relevant in this context, since they
characterize capture and storage of \chem{CO_2},
i.e.\ phase separation and phase equilibrium.
The present study concentrates on understanding homogeneous
nucleation and VLE properties for quaternary mixtures consisting of \chem{CO_2}, nitrogen (\chem{N_2}),
oxygen (\chem{O_2}), and argon (\chem{Ar}) by means of molecular simulation. Thereby,
the properties of the investigated fluid itself can be fully isolated from phenomena
induced by impurities or boundary effects in the vicinity of a solid wall,
which would be much harder to accomplish in an experimental arrangement.

Although a systematic discussion of \chem{CO_2} nucleation
was, surprisingly enough, published for the ambient conditions
prevailing on Mars \citep{MVLMKSK05}, to the authors' knowledge no
analogous study is available for the ecologically and technically more relevant
atmosphere composition of our own planet.
The present work closes this gap and characterizes the air pressure
effect on the condensation process in a \vapour{} containing more
\chem{CO_2} than at saturation on the basis of molecular models that are known to
accurately reproduce VLE properties.

Direct MD simulation of nucleation in the canonical ensemble with the
Yasuoka-Matsumoto (YM) method is an established approach \citep{YM98, MKEY07}.
It can be successfully applied to the regime where the supersaturation
is sufficiently high to permit significant \nucleus{} formation in a nanoscopic volume
within a few nanoseconds, but not as high that the nucleation rate $\nuclrate$
is affected by depletion of the \vapour{} due to \nucleus{} formation
within the same time interval \citep{HVBGRWSH08, HVH08, CWSWR09}. Lower nucleation
rates can be determined by forward flux sampling or methods based on
umbrella sampling \citep{VSF05}, whereas in the immediate vicinity of the
spinodal line, where the metastable state would otherwise break down,
a stationary value for the nucleation rate can be obtained by grand canonical
MD simulation with McDonald's daemon \citep{HV09}.

The present work focuses on the intermediate regime, where the YM method
is viable, and compares simulation results with theoretical predictions
on the basis of two variants of the classical nucleation theory (CNT).

%T% 
%T% \citep{HVH08}  % simulation of single droplets in equilibrium
%T% \citep{Zhukovitskii95, HL08}  % ... and in non-equilibrium
%T% 

%%%%%%%%%%%%%%%%%%%%%%%%%%%%%%%%%%%%%%
% hier beginnt der theoretische Teil %
%%%%%%%%%%%%%%%%%%%%%%%%%%%%%%%%%%%%%%

\section{The pressure effect}
\label{sec:wedekind}

If the pressure effect (\txtCNT{}) is taken into account, the free energy $\gibbs$ of
formation according to CNT is given by \citep{WHBR08}
\begin{equation}
   \CNT{\differential\gibbs} =
      \left(\sat\chempot - \chempot - \frac{\sat\partialpressure - \pressure}{\liq\density}\right)
         \differential\nuclsize + \surfacetension \differential\area,
\end{equation}
for a \nucleus{} containing $\nuclsize$ molecules,
whereas the more commonly used -- but thermodynamically inconsistent -- simplified
classical (\txtCNTnop{}) variant of CNT implies
\begin{equation}
   \CNTnop{\differential\gibbs} =
      \left(\sat\chempot - \chempot\right) \differential\nuclsize + \surfacetension \differential\area.
\end{equation}
Therein, $\sat\chempot$ and $\sat\partialpressure$ as well as $\chempot$ and $\pressure$
represent the saturated and supersaturated values of chemical potential and pressure,
respectively. While $\sat\partialpressure$ refers to the pure substance \vapor{}
pressure of the nucleating component, $\pressure$ is the total pressure including the
pressure contribution of an inert carrier gas, if present. 
The \nucleus{} surface tension is given by $\surfacetension$, the area of the
surface of tension is $\area$, and $\liq\density$ represents the density of the
liquid which is assumed to be incompressible.
Where the chemical potential difference $\Deltamu = \chempot - \sat\chempot$ appears in the SC
expression, PE applies an \qq{effective} difference \citep{WHBR08}
\begin{equation}
\Deltamueff = \Deltamu + \frac{\sat\partialpressure - \pressure}{\liq\density},
\label{eqn:Deltamueff}
\end{equation}
which is always smaller than $\Deltamu$. Assuming a spherical shape for the emerging \droplets, i.e.\
\begin{equation}
   \differential\area = \left(\frac{32\pi}{3{\liq\density}^2\nuclsize}\right)^{1\slash{}3} \differential\nuclsize,
\end{equation}
and neglecting the curvature dependence of the surface tension, the free energy barrier
of the nucleation process according to the PE variant of CNT
\begin{equation}
   \DGcritCNT = \frac{16\pi\planartension^3}{3(\liq\density\Deltamueff)^2}
      + \Deltamueff - \pi^{1\slash{}3}\left(\frac{6}{\liq\density}\right)^{2\slash{}3}\planartension,
\label{eqn:DGcrit}
\end{equation}
is reached for the critical \nucleus{} size
\begin{equation}
   \NcritCNT = \frac{32\pi}{3{\liq\density}^2}
      \left(\frac{\planartension}{\Deltamueff}\right)^3.
\label{eqn:criticalnuclsize}
\end{equation}
These expressions use the surface tension $\planartension$
of the planar \vapour-liquid interface as well as the \qq{internally consistent}
approach of \cite{BK72} which equates single-molecule \droplets{} with
\vapour{} monomers and assigns them a free energy of formation $\Delta\gibbs = 0$.
The same results follow for the SC variant, with $\Deltamu$ instead
of $\Deltamueff$ in Eqs.\ (\ref{eqn:DGcrit}) and (\ref{eqn:criticalnuclsize}).

The long-term growth probability $\committor(\threshold)$ of a \droplet{}
containing $\threshold$ molecules can be given by 
\begin{equation}
   \committor(\threshold) = \int_1^\threshold
      \exp\left(\frac{2\Delta\gibbs(\nuclsize)}{\kboltz\temperature}\right)
         \differential\nuclsize \, \left[\int_1^\infty
            \exp\left(\frac{2\Delta\gibbs(\nuclsize)}{\kboltz\temperature
               }\right) \differential\nuclsize\right]^{-1},
\label{eqn:committor}
\end{equation}
under the approximation that
the reaction coordinate of the nucleation process only depends on the \droplet{} size
order parameter $\nuclsize$ \citep{HMV09}.
The transition rate of \vapour{} monomers of the nucleating component
through an interface, normalized by the surface area of the interface, is 
\begin{equation}
   \trate = \partialpressure\left(2\pi\supmass\kboltz\temperature\right)^{-1\slash{}2},
\end{equation}
according to kinetic gas theory,
where $\kboltz$ is the Boltzmann constant, $\supmass$ is the molecular mass
and $\partialpressure$ the pressure of the nucleating component in
its pure gaseous state at the same partial density.
%T%
%T% % The critical nucleus is equivalently characterized by T = E
%T% \citep{SS98}  % emission rate = transition rate
%T%
The isothermal nucleation rate according to CNT is \citep{FRLP66}
\begin{equation}
   \nuclrateiso = \zeldovich\critical{\area}\trate\densitymon
      \exp\left(\frac{\critical{-\Delta\gibbs}}{\kboltz\temperature}\right),
\label{eqn:Jiso}
\end{equation}
which depends on the surface area $\critical{\area}$ of the critical \droplet,
the \cite{Zeldovich42} factor 
\begin{equation}
   \zeldovich = \left(\frac{-\differential^2\gibbs\slash\differential\nuclsize^2
      }{2\pi\supmass\kboltz\temperature}
         \right)_{\nuclsize = \critical{\nuclsize}}^{1\slash{}2},
\end{equation}
as well as the density of \vapour{} monomers
belonging to the nucleating component $\densitymon$.
\Droplet{} overheating due to rapid growth, however, is neglected
in the expression for $\nuclrateiso$. The monomer
density in the metastable state can be obtained from
\begin{equation}
   \density\supfraction = \densitymon \sum_{\nuclsize = 1}^{\critical\nuclsize} \nuclsize
      \exp\left(\frac{-\Delta\gibbs(\nuclsize)}{\kboltz\temperature}\right),
\end{equation}
a series that usually converges very fast, wherein $\density$ is the total
density of the supersaturated \vapour{} and $\supfraction$ is the
mole fraction of the nucleating component in the supersaturated \vapour.
The presence of a carrier gas also influences the thermalization of
growing \nuclei{}, facilitating the heat transfer
from the liquid to the surrounding \vapour{} and thereby decreasing the
amount of overheating. This effect is covered by the thermal
non-accomodation prefactor of \cite{FRLP66}
\begin{equation}
   \nuclrate = \frac{\federb^2}{\federb^2 + \federq^2}\nuclrateiso,
\end{equation}
consisting of
\begin{equation}
   \federq = \vap\enthalpy - \liq\enthalpy - \frac{\kboltz\temperature}{2}
      - \surfacetension\left(\frac{\differential\area}{\differential\nuclsize
         }\right)_{\nuclsize = \critical\nuclsize},
\end{equation}
as well as the mean square fluctuation of the kinetic energy
for the \vapour{} molecules \citep{FRLP66, WHBR08}
\begin{equation}
   \federb^2 = \kboltz\temperature^2 \left(\isochoric{0} + \frac{\kboltz}{2}\right) 
      \sum_{\component = 0}^{\numcomponents} \frac{
         \molarfraction{\component}\mass_0^{1\slash{}2}(\isochoric{\component} + \kboltz\slash{}2)
            }{\molarfraction{0}\mass_\component^{1\slash{}2}(\isochoric{0} + \kboltz\slash{}2)},
\end{equation}
where $\liq\enthalpy$ and $\vap\enthalpy$ are saturated liquid and \vapor{} enthalpy, respectively,
$\isochoric{\component}$ is the isochoric heat capacity
of component $\component$, where $\component = 0$ indicates
the nucleating component and $1 \leq \component \leq \numcomponents$ the components
of the inert carrier gas. Within the scope of the present study, the heat capacity of the pure
saturated \vapour{} is used for $\isochoric{0}$ whereas for the other $\isochoric{\component}$
the value in the limit of infinite dilution is used, since CNT assumes
the carrier gas to have ideal properties \citep{WHBR08}.

Within the framework of CNT, it follows that the prefactor
\begin{equation}
   \zeldovich\critical{\area} = \frac{2}{\liq\density}
      \left(\frac{\planartension}{\kboltz\temperature}\right)^{1\slash{}2},
\end{equation}
does not depend on the pressure contribution of the carrier gas, and neither does $\trate$,
while the influence of the carrier gas pressure on $\densitymon$, which is usually
similar in magnitude to $\density\supfraction$, is of minor significance. This eliminates all
contributions to the pressure effect except for those discussed by \cite{WHBR08} as
\begin{equation}
   \frac{\CNT\nuclrate}{\CNTnopnuclrateiso} = \frac{\CNTfederbb}{\CNTfederbb + \CNTfederqq}
      \, \frac{\CNTnuclrateiso}{\CNTnopnuclrateiso}.
\label{eqn:origW}
\end{equation}
Normalized to unity for the pure
fluid ($\supfraction = 1$) with the PE variant of CNT,
the \cite{WHBR08} pressure effect can be expressed as
\begin{equation}
   \Weffect(\supfraction) % = \frac{\CNT\nuclrate(\supfraction)}{\CNT\nuclrate(\supfraction = 1)}
      = \frac{\CNTfederbb(\supfraction) \, \left[\CNTfederbb(\supfraction = 1) + \CNTfederqq(\supfraction = 1)\right]}{\left[\CNTfederbb(\supfraction) + \CNTfederqq(\supfraction)\right] \, \CNTfederbb(\supfraction = 1)} \,
            \exp\left(\frac{\DGcritCNT(\supfraction = 1) - \DGcritCNT(\supfraction)}{\kboltz\temperature}\right),
\end{equation}
given that $\CNTnopnuclrateiso$, the denominator of Eq.\ (\ref{eqn:origW}), does not
depend on $\supfraction$.

Under certain conditions, the pressure effect does not exceed the experimental uncertainty and
can thus be neglected \citep{IWWWS04}. In other cases, however, the influence can
be experimentally detected, with apparently contradictory results: sometimes
$\nuclrate$ increases with the amount of carrier gas, in other
cases the opposite tendency is observed \citep{HBZSKVL06, BHWVKZSL08}.
The $\Weffect$ factor explains this in principle, since it combines the thermal non-accomodation
factor, which increases with $\supfraction \to 0$, and the free energy effect that leads
to an effective chemical potential difference $\Deltamueff < \Deltamu$.
Depending on the thermodynamic conditions,
each of these contributions can be predominant \citep{WHBR08}.

The main inaccuracies of CNT concern the surface tension as well as
the \droplet{} surface area.
For the surface tension, deviations from the capillarity approximation
$\surfacetension \approx \planartension$ are known to occur for nanoscopically
curved interfaces \citep{MA03, SU03, HVH08}.
The \cite{Tolman49} approach implies
huge deviations from $\planartension$ for \droplets{} on the molecular
length scale corresponding to the Tolman length $\tolmanlength$. 
In particular, for $\radius > \tolmanlength$ the dependence of $\surfacetension$ on the
surface of tension radius $\radius$ can be expressed as
\begin{equation}
   \surfacetension = \planartension \exp\left[-2\left(\frac{\tolmanlength}{\radius}\right) + \left(\frac{\tolmanlength}{\radius}\right)^2 - \frac{2}{9}\left(\frac{\tolmanlength}{\radius}\right)^3 + \landau{\left[\tolmanlength\slash\radius\right]^4}\right],
\label{eqn:corr1}
\end{equation}
whereas $\surfacetension \sim \radius$ becomes valid for $\radius \ll \tolmanlength$.
From the Laplace equation along with the \cite{Tolman49} approach it can
be deduced that the area of the surface of tension is given by
\begin{equation}
   \differential\area = \frac{2\differential\nuclsize}{\radius\liq\density},
\label{eqn:corr2}
\end{equation}
for an incompressible fluid, where the relation
\begin{equation}
   \radius = \left(\frac{3\nuclsize}{4\pi\liq\density}\right)^{1\slash{}3} - \tolmanlength,
\end{equation}
follows for the dependence of the surface of tension radius on $\nuclsize$.
Overall, this leads to a significantly increased surface area for
\droplets{} on the molecular length scale.

\section{Simulation methods and models}
\label{sec:sim}

The evaluation of the theoretical predictions relies on knowledge about the
chemical potential difference between the saturated and the supersaturated
state. This was obtained for pure \chem{CO_2} by Gibbs-Duhem integration
\begin{equation}
   \Deltamu(\temperature, \pressure) =
      \int_{\sat\pressure(\temperature)}^\pressure \frac{\differential\pressure}{\density(\temperature, \pressure)},
\end{equation}
using MD simulation results of small systems ($\absnum \approx 10$ $000$)
in the metastable \vapour{} regime.
The carrier gas influence according to the presented variants of CNT was evaluated
by assuming ideal gas properties for air as well as ideal mixing behavi\oou{}r, i.e.\
\begin{equation}
   \pressure = (1 - \supfraction)\density\kboltz\temperature + \partialpressure.
\end{equation}
For the homogeneous nucleation simulations, the YM method was applied to relatively
large, but still nanoscopic systems with $\absnum(\chem{CO_2}) = 300$ $000$.
The total number of molecules was up to $\absnum = 900$ $000$ such that the
carrier gas with
$\absnum(\chem{N_2}) : \absnum(\chem{O_2}) : \absnum(\chem{Ar}) = 7812 : 2095 : 93$
corresponded to the earth's atmosphere composition.
The condensation process is thereby regarded as a succession of three characteristic
stages: relaxation, nucleation, and \nucleus{} growth \citep{CP07}. During the nucleation stage,
the droplet formation rate $\formationrate(\threshold)$, i.e.\ the number of \nuclei{}
containing at least $\threshold$ molecules formed over time,
is approximately constant \citep{YM98}. The droplet
formation rate depends on the threshold size $\threshold$ and is related to
the nucleation rate by
\begin{equation}
   \frac{\formationrate(\threshold)}{\volume} = \frac{\nuclrate}{\committor(\threshold)},
\label{eqn:formationrate}
\end{equation}
since $\committor(\threshold)$ indicates the probability for a \nucleus{} containing
$\threshold$ molecules to reach macroscopic size.
Liquid and \vapour{} were distinguished according to a \cite{Stillinger63} criterion
such that molecules separated by distances of their centres of mass below $5.08$ \AA{}
were considered as part of the liquid.
Biconnected components, where any single
connection can be eliminated without disrupting the internal connectivity, were defined
to be liquid \nuclei.

%%%%%%%%%%%%%%%%%%%%%%%%%%%%%%%%%%%%%%%%%%%%%%%%%%%%%%
% hier beginnt der Abschnitt zu den Modellen und VLE %
%%%%%%%%%%%%%%%%%%%%%%%%%%%%%%%%%%%%%%%%%%%%%%%%%%%%%%

\begin{figure}[h!]
\centering
\includegraphics[width=8.4cm]{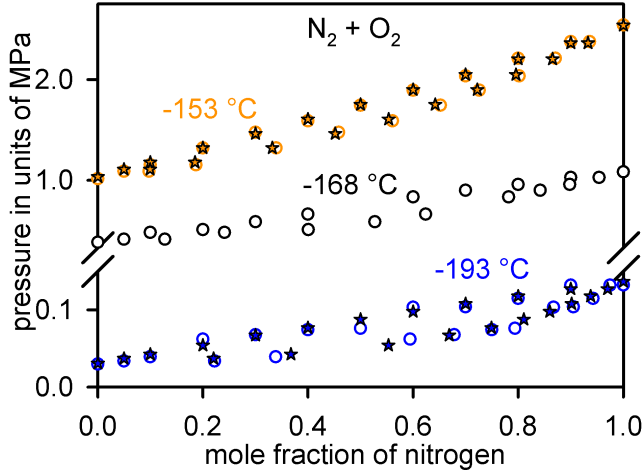}
\caption{
   (Col\oou{}r on the web, b/w in print.)
   Experimental data ($\star$) of \cite{Dodge27} and simulation results ($\circ$) of \cite{VHH09},
   using the Grand Equilibrium method with the molecular models given in Tab.\ \ref{tabmod}, for
   VLE of binary mixtures containing nitrogen and oxygen at temperatures of
   $\temperature$ = $-193$, $-168$, and $-153$ $^\circ$C.
}
\label{figbin}
\end{figure}

Molecular models for \chem{Ar}, which can be represented by one
Lennard-Jones (LJ) site, and for \chem{CO_2}, \chem{N_2} as well as \chem{O_2}, which can
be represented by two LJ sites separated by the elongation $\elongation$
with a superimposed quadrupole moment $\quadrupole$ in the molecule's
centre of mass (2CLJQ), were adjusted to pure fluid VLE data by \cite{VSH01}, cf.\ Tab.\ \ref{tabmod}.
If adequate values for the unlike dispersive interaction energy are used, so that
binary VLE are reproduced correctly \citep{VHH09}, cf.\ Fig.\ \ref{figbin},
ternary mixtures are accurately described without any further adjustment \citep{HVH09}.
In Tab.\ \ref{tabmod}, the unlike energy parameters are indicated
in terms of the binary interaction parameter $\unlike$ of the
modified \cite{Berthelot98} combining rule \citep{SVH07}
\begin{equation}
   \LJenergy_{\fluidA\fluidB} = \unlike(\LJenergy_\fluidA \LJenergy_\fluidB)^{1\slash{}2},
\end{equation}
while  the unlike LJ size parameter is determined as an arithmetic mean according to
the \cite{Lorentz81} combining rule.
This approach has also been validated with an emphasis on \chem{CO_2} in particular,
confirming its viability for mixtures with \chem{N_2} and \chem{O_2} \citep{VKBMH09} as well
as hydrogen bonding fluids \citep{SSVH07}.

On that basis, quaternary phase equilibria were determined using the Grand Equilibrium
method \citep{VH02}, introducing \chem{Ar} into the the system studied by \cite{VKBMH09}.
The Grand Equilibrium method calculates the \vapour{} pressure
$\sat{\pressure}$ as well as all dew line mole fractions $\vapfraction_\component$
by simulation for a specified temperature $\temperature$ and a specified
bubble line mole fraction $\liqfraction_\component$ for all components of the mixture.
Although no experimental data are available for the quaternary mixture, the simulation
results can be trusted due to the extensive validation of the models with respect
to the VLE behavi\oou{}r for all of the six binary \citep{VHH09} and two of the
four ternary subsystems \citep{HVH09}, i.e.\ \chem{N_2} + \chem{O_2} + \chem{Ar}
as well as \chem{CO_2} + \chem{N_2} + \chem{O_2}.

%% Table 1(t): molecular models and unlike interaction parameters
%
\begin{table}[t!]
\centering
\begin{tabular}{l|lcccc|cccc}
 & type
 & $\LJlength$ [\AA] & $\LJenergy$ [meV] & $\elongation$ [\AA] & $\quadrupole$ [$\echarge$\AA{}$^2$]
                     & $\unlike\left(\chem{Ar}\right)$ & $\unlike\left(\chem{O_2}\right)$
                     & $\unlike\left(\chem{N_2}\right)$ \\ \hline
\chem{CO_2} & 2CLJQ & $2.9847$ & $11.394\phantom{0}$ & $2.418\phantom{0}$ & $0.78985$
            & $0.999$ & $0.979$ & $1.041$ \\
\chem{N_2}  & 2CLJQ & $3.3211$ & $\phantom{0}3.0072$ & $1.046\phantom{0}$ & $0.29974$
            & $1.008$ & $1.007$ & \\
\chem{O_2}  & 2CLJQ & $3.1062$ & $\phantom{0}3.7212$ & $0.9699$ & $0.16824$
            & $0.988$ & & \\
\chem{Ar}   & LJ    & $3.3967$ & $10.087\phantom{0}$ & & & & & \\
\hline
\end{tabular}
\caption{
   \label{tabmod}
   Molecular model parameters of \cite{VSH01} and binary interaction parameters $\unlike$ adjusted to binary VLE data \citep{VHH09}.
}
\end{table}

%% Figure 1(h): binary mixtures (reproduction)
%

%T% 
%T% % Peng-Robinson
%T% %
%T% \citep{BVFM08, VKBMH09}
%T% 

%T% 
%T% % Henry model for the ternary mixture CO2+N2+O2
%T% %
%T% \citep{VKBMH09}
%T% 

\section{Simulation results}
\label{sec:res}

%T% 
%T% % comparison with Peng-Robinson
%T% 

Grand Equilibrium simulations of the quaternary
mixture $\chem{CO_2}$ + $\chem{N_2}$ + $\chem{O_2}$ + $\chem{Ar}$ were conducted
for VLE covering a broad temperature range with $\chem{CO_2}$ bubble line mole fractions
$\liqfraction(\chem{CO_2})$ of $0.910$, $0.941$, and $0.969$, cf.\ Tab.\ \ref{tabqtr}. Except for
the highest temperature, which corresponds to $93$ $\%$ of the critical temperature $\Tcrit$
for pure \chem{CO_2}, 
the mole fractions $\vapfraction(\chem{N_2})$, $\vapfraction(\chem{O_2})$,
and $\vapfraction(\chem{Ar})$ on the dew line are one order of magnitude higher
than the corresponding bubble line mole fractions. This confirms that for
temperatures sufficiently below $\Tcrit(\chem{CO_2})$, air only accumulates to a
limited extent in the liquid phase. As a first approximation, it can therefore
be treated as a carrier gas for \chem{CO_2} nucleation so that the PE variant
of CNT with a single nucleating component can be applied,
as opposed to more complex mixtures such as ethanol + hexanol \citep{SV93},
water-alcohol mixtures \citep{VSLK94, SVW95},
or water + nonane + butanol \citep{NCS07}, where multi-component nucleation
occurs.

%% Table 2(b): quaternary mixture
%
\begin{table}[b!]
\centering
\begin{tabular}{ll|lllllll}
$\temperature$ [$^\circ$C] & $\liqfraction\left(\chem{CO_2}\right)$ & $\sat\pressure$ [MPa]
                           & $\vapfraction\slash\liqfraction\left(\chem{N_2}\right)$ & $\vapfraction\slash\liqfraction\left(\chem{O_2}\right)$
                           & $\vapfraction\slash\liqfraction\left(\chem{Ar}\right)$
                           & $\liq\density$ [mol/l] & $\vap\density$ [mol/l]
                           & $\vap\enthalpy - \liq\enthalpy$ [kJ/mol] \\ \hline
$-90.3$ & $0.969$
      & $2.53(8)$ & $40(2)$ & $24.5(9)$ & $24.7(9)$ 
      & $29.24(1)$ & $1.854(2)$ & $17.092(9)$ \\
      & $0.941$
      & $3.9(1)$ & $21.5(8)$ & $13.9(5)$ & $13.7(5)$
      & $29.13(1)$ & $3.031(5)$ & $16.27(1)$ \\
      & $0.910$
      & $6.0(2)$ & $13.7(4)$ & $\phantom{0}9.1(3)$ & $\phantom{0}9.0(3)$
      & $29.01(2)$ & $5.29(2)$ & $15.12(1)$ \\
\hline
$-40.3$ & $0.941$
      & $4.38(5)$ & $14.0(2)$ & $10.8(1)$ & $10.3(1)$
      & $24.94(3)$ & $2.663(5)$ & $13.30(1)$ \\
      & $0.910$
      & $5.30(4)$ & $\phantom{0}9.7(1)$ & $\phantom{0}7.72(8)$ & $\phantom{0}7.34(8)$
      & $24.47(2)$ & $3.247(7)$ & $12.64(1)$ \\
\hline
$\phantom{0-}9.7$ & $0.969$
                & $5.97(4)$ & $\phantom{0}5.6(1)$ & $\phantom{0}4.88(8)$ & $\phantom{0}4.72(7)$
                & $19.3(2)$ & $3.90(1)$ & $\phantom{0}8.55(5)$ \\
                & $0.941$
                & $6.98(3)$ & $\phantom{0}4.31(5)$ & $\phantom{0}3.82(5)$ & $\phantom{0}3.67(4)$
                & $18.55(9)$ & $4.63(2)$ & $\phantom{0}7.79(4)$ \\
\hline
\end{tabular}
\caption{
   \label{tabqtr}
   VLE data for the quaternary
   system $\chem{CO_2}$ + $\chem{N_2}$ + $\chem{O_2}$ + $\chem{Ar}$.
   The liquid composition is equimolar in nitrogen, oxygen, and argon, i.e.\ $\liqfraction(\chem{N_2})
   = \liqfraction(\chem{O_2}) = \liqfraction(\chem{Ar}) = \left[1 - \liqfraction(\chem{CO_2})\right]$
   $\slash$ $3$, and values in parentheses indicate the uncertainty in terms of the last given digit.
}
\end{table}

%T% 
%T% % Erweiterung des Henry-Modells
%T% 

From MD simulation of small metastable systems, the spinodal value $\sps$ of the
supersaturation with respect to density, which is defined as
\begin{equation}
   \supersatrho = \frac{\density\supfraction}{\vap\density(\temperature)},
\end{equation}
wherein $\vap\density(\temperature)$ is the saturated \vapour{} density of pure $\chem{CO_2}$,
was determined to be in the range $4.3 \leq \sps \leq 5.1$
at $-44.8$ $^\circ$C, $3.6 \leq \sps \leq 4.3$ at $-34.8^\circ$C, and $3.0 \leq \sps \leq 3.6$
at $-23$ $^\circ$C for $\supfraction = 1$.
At these temperatures, canonical ensemble MD simulations for $\chem{CO_2}$
nucleation were conducted using the YM method with $\chem{CO_2}$ mole fractions of
$1\slash{}3$, $1\slash{}2$, and $1$ at supersaturations below the spinodal value $\sps$, but
still high enough to obtain statistically reliable \nucleus{} formation rates in a
nanoscopic volume on the timescale of a few nanoseconds.

\begin{figure}[h!]
\centering
\includegraphics[width=8.4cm]{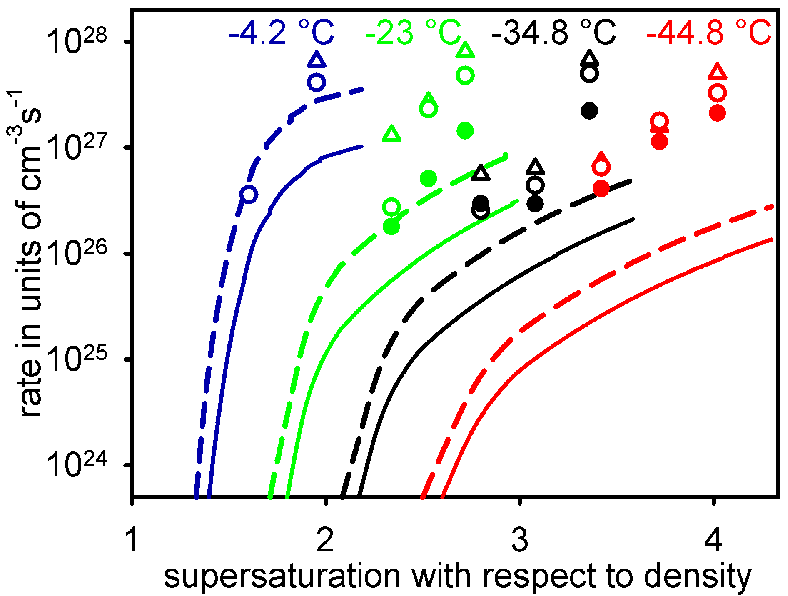}
\caption{
   (Col\oou{}r on the web and in print.)
   Pure \chem{CO_2} nucleation rate $\nuclrate$ according to the PE (---) and SC (-- --)
   variants of CNT in comparison to $\formationrate\slash\volume$ for threshold
   sizes of $\threshold$ $=$ ($\Delta$) $50$, ($\circ$) $75$, and $250$ ($\bullet$)
   molecules from canonical ensemble MD simulation over
   supersaturation $\supersatrho = \density\slash\vap\density(\temperature)$
   at temperatures of $\temperature$ $=$ $-44.8$, $-34.8$,
   $-23$, and $-4.2$ $^\circ$C. The simulation results for $\temperature
   = -4.2$ $^\circ$C are taken from previous work and were obtained using a
   different cluster criterion \citep{HVBGRWSH08}.
}
\label{figpur}
\end{figure}

YM \droplet{} formation rates $\formationrate$
from the present work as well
as a previous study \citep{HVBGRWSH08} are shown in Fig.\ \ref{figpur}
for $\supfraction = 1$, i.e.\ pure \chem{CO_2}. The dependence of
$\formationrate$ on the threshold size $\threshold$ reproduces the typical
picture: for low threshold sizes (probably smaller than $\critical\nuclsize$)
the droplet formation rate can be elevated by several orders of magnitude,
and it converges for $\threshold \gg \critical\nuclsize$ under the condition
that the depletion of the \vapour{} can be neglected \citep{YM98}. For very large
values of $\threshold$ -- not shown in Fig.\ \ref{figpur} -- the \droplet{} formation
rate decreases again, because the presence of many large \droplets{} implies
that a substantial amount of the \vapour{} monomers have already been
consumed by the emerging liquid phase.

The most striking observation is that while both variants of CNT
predict the value of $\nuclrate$ in the spinodal limit to increase
with temperature -- mainly because $\temperature$ occurs in the
denominator of the exponential in the Arrhenius term of Eq.\ (\ref{eqn:Jiso}) -- the
simulation results do not exhibit any significant temperature dependence for
the attainable value of $\nuclrate$. In the spinodal limit, the nucleation rate
appears to be about $\nuclrate(\temperature, \sps) \approx 10^{27}$ cm$^{-3}$s$^{-1}$
over the whole temperature range.
The pressure effect in the pure fluid, expressed
by $\CNTnop\nuclrate \slash \CNT\nuclrate$, is most significant at
high temperatures, because this corresponds to a lower density of the liquid
and because $\Deltamu$ is smaller so that the relative deviation between
$\Deltamueff$ and $\Deltamu$ is increased.

%% Figure 2(h): pure carbon dioxide nucleation
%

Table \ref{tabnuc1} indicates the results for the carrier gas effect on
\chem{CO_2} nucleation at $\temperature = -44.8$ $^\circ$C. As usual,
$\formationrate$ decreases when larger values of the
threshold size $\threshold$ are regarded, but it can also be seen
that this effect is clearly stronger when more air is present in the system.
This leads to values for the overall carrier gas effect $\gaseffect$ on the
\droplet{} formation rate
\begin{equation}
   \gaseffect = \frac{\formationrate(\supersatrho, \supfraction, \threshold)
      }{\formationrate(\supersatrho, \supfraction = 1, \threshold)},
\end{equation}
that are greater than unity for relatively small values of $\threshold$,
but converge to values significantly below unity as $\threshold$ is increased.
This result can be understood if the carrier gas effect
on $\critical\nuclsize$ according to Eqs.\ (\ref{eqn:Deltamueff}) and
(\ref{eqn:criticalnuclsize}) is considered: with a higher total pressure,
$\Deltamueff$ decreases which affects the critical \droplet{} size to
the third power, leading to significantly larger values of $\NcritCNT$,
cf.\ Tab.\ \ref{tabnuc1}. Thus, for relatively small threshold sizes,
the long-term growth probability is significantly smaller, cf.\
Eq.\ (\ref{eqn:committor}), which in turn increases the \droplet{}
formation rate according to Eq.\ (\ref{eqn:formationrate}).
Hence, the apparently contradictory values of $\gaseffect$ are
actually consistent and correspond to a negative dependence of
the nucleation rate on the carrier gas density. 

%% Table 3a(t): all results for J_i at T = 228.35 K
%
\begin{table}[t!]  
\centering
\begin{tabular}{ccc|cc|clcc}
$\supersatrho$ & $\supfraction$ & $\threshold$
               & $\formationrate\slash\volume$ [cm$^{-3}$s$^{-1}$]
               & $\gaseffect$ & $\NcritCNT$ & $\Weffect$ 
               & $\CNT\nuclrate$ [cm$^{-3}$s$^{-1}$]
               & $\CNT{\nuclrate\slash\committor}$ [cm$^{-3}$s$^{-1}$] \\ \hline
$3.42$ & $1/3$ & $\phantom{0}50$
     & $2.7 \mult 10^{27}$ & $3.6\phantom{0}$
     & $61$ & $8.5\mult10^{-3}$
     & $5.2 \mult 10^{23}$ & $3.0 \mult 10^{24}$ \\
     &     & $\phantom{0}75$
     & $5.2 \mult 10^{26}$ & $0.79$
     & & 
     & & $6.1 \mult 10^{23}$ \\
     &     & $\phantom{0}85$
     & $3.8 \mult 10^{26}$ & $0.63$
     & & 
     & & $5.4 \mult 10^{23}$ \\
     & $1/2$ & $\phantom{0}50$
     & $1.1 \mult 10^{27}$ & $1.5\phantom{0}$
     & $44$ & $0.17$
     & $9.8 \mult 10^{24}$ & $1.4 \mult 10^{25}$ \\
     &     & $\phantom{0}75$
     & $3.1 \mult 10^{26}$ & $0.47$
     & & 
     & & $9.8 \mult 10^{24}$ \\
     &     & $\phantom{0}85$
     & $2.4 \mult 10^{26}$ & $0.39$
     & & 
     & & $9.8 \mult 10^{24}$ \\
     & $1\phantom{/1}$ & $\phantom{0}50$
     & $7.3 \mult 10^{26}$ & $1\phantom{.00}$
     & $33$ & $1$
     & $5.0 \mult 10^{25}$ & $5.2 \mult 10^{25}$ \\
     &               & $\phantom{0}75$
     & $6.5 \mult 10^{26}$ & $1\phantom{.00}$
     & & 
     & & $5.0 \mult 10^{25}$ \\
     &               & $\phantom{0}85$
     & $6.1 \mult 10^{26}$ & $1\phantom{.00}$
     & & 
     & & $5.0 \mult 10^{25}$ \\ \hline
$3.72$ & $1/3$ & $\phantom{0}50$
     & $1.4 \mult 10^{28}$ & $8.7\phantom{0}$
     & $61$ & $5.8\mult10^{-3}$
     & $5.9 \mult 10^{23}$ & $3.4 \mult 10^{24}$ \\
     &     & $\phantom{0}85$
     & $5.0 \mult 10^{27}$ & $2.9\phantom{0}$
     & & 
     & & $6.1 \mult 10^{23}$ \\
     &     & $150$        
     & $7.0 \mult 10^{26}$ & $0.63$
     & & 
     & & $5.9 \mult 10^{23}$ \\
     & $1/2$ & $\phantom{0}50$
     & $2.3 \mult 10^{27}$ & $1.4\phantom{0}$
     & $43$ & $0.15$
     & $1.3 \mult 10^{25}$ & $1.8 \mult 10^{25}$ \\
     &     & $\phantom{0}85$
     & $1.2 \mult 10^{27}$ & $0.71$
     & & 
     & & $1.3 \mult 10^{25}$ \\
     &     & $150$       
     & $4.4 \mult 10^{26}$ & $0.38$
     & & 
     & & $1.3 \mult 10^{25}$ \\
     & $1\phantom{/1}$ & $\phantom{0}50$
     & $1.6 \mult 10^{27}$ & $1\phantom{.00}$
     & $32$ & $1$ 
     & $7.9 \mult 10^{25}$ & $8.1 \mult 10^{25}$ \\
     &               & $\phantom{0}85$
     & $1.7 \mult 10^{27}$ & $1\phantom{.00}$
     & & 
     & & $7.9 \mult 10^{25}$ \\
     &               & $150$        
     & $1.1 \mult 10^{27}$ & $1\phantom{.00}$
     & & 
     & & $7.9 \mult 10^{25}$ \\ \hline
$4.02$ & $1/3$ & $\phantom{0}85$
     & $7.7 \mult 10^{27}$ & $2.5\phantom{0}$
     & $61$ & $3.8\mult10^{-3}$
     & $5.4 \mult 10^{23}$ & $5.6 \mult 10^{23}$ \\
     &     & $150$      
     & $3.4 \mult 10^{27}$ & $1.6\phantom{0}$
     & & 
     & & $5.4 \mult 10^{23}$ \\
     &     & $300$
     & $6.3 \mult 10^{26}$ & $0.47$
     & & 
     & & $5.4 \mult 10^{23}$ \\
     & $1/2$ & $\phantom{0}85$
     & $2.4 \mult 10^{27}$ & $0.77$
     & $42$ & $0.13$
     & $1.7 \mult 10^{25}$ & $1.7 \mult 10^{25}$ \\
     &     & $150$          
     & $1.1 \mult 10^{27}$ & $0.49$
     & & 
     & & $1.7 \mult 10^{25}$ \\
     &     & $300$        
     & $6.1 \mult 10^{26}$ & $0.46$
     & & 
     & & $1.7 \mult 10^{25}$ \\
     & $1\phantom{/1}$ & $\phantom{0}85$
     & $3.2 \mult 10^{27}$ & $1\phantom{.00}$
     & $31$ & $1$
     & $1.1 \mult 10^{26}$ & $1.1 \mult 10^{26}$ \\
     &               & $150$          
     & $2.1 \mult 10^{27}$ & $1\phantom{.00}$
     & & 
     & & $1.1 \mult 10^{26}$ \\
     &               & $300$         
     & $1.3 \mult 10^{27}$ & $1\phantom{.00}$
     & & 
     & & $1.1 \mult 10^{26}$ \\ \hline
\end{tabular}
\caption{
   \label{tabnuc1}
   \Droplet{} formation rate and carrier gas effect from YM canonical
   ensemble MD simulation as well as critical \droplet{} size (in
   molecules), normalized \cite{WHBR08} pressure effect, nucleation
   rate and \droplet{} formation rate according to the PE variant of CNT,
   in dependence of supersaturation and mole fraction
   of $\chem{CO_2}$ in the \vapour{} as well as the YM threshold size (in
   molecules) at a temperature of $\temperature$ = $-44.8$ $^\circ$C.
}
\end{table}

\begin{figure}[h!]
\centering
\includegraphics[width=8.4cm]{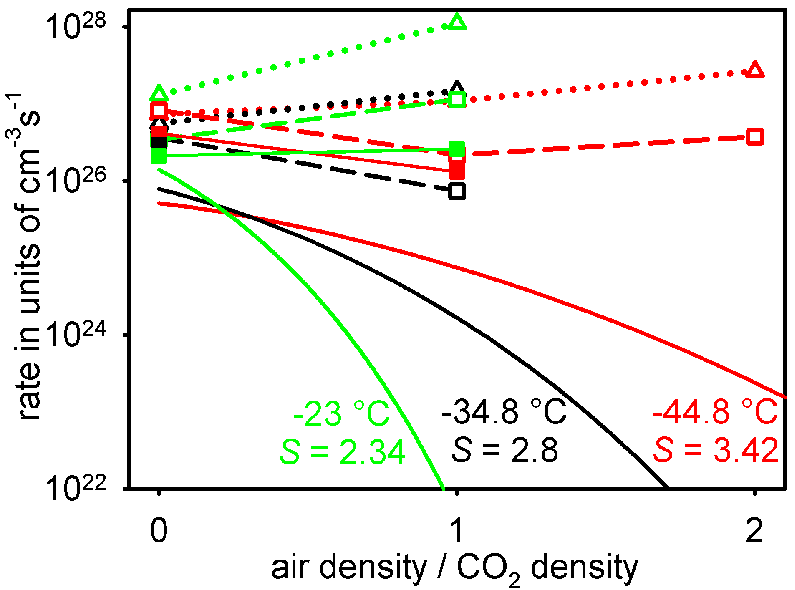}
\caption{
   (Col\oou{}r on the web and in print.)
   Nucleation rate $\CNT\nuclrate$ according to CNT with the pressure
   effect in comparison to $\formationrate\slash\volume$ for threshold
   sizes of $\threshold$ = $50$ ($\Delta$ $\cdot$ $\cdot$ $\cdot$ $\Delta$),
   $100$ ($\square$ -- -- $\square$), and $150$ ($\blacksquare$ --- $\blacksquare$)
   molecules from canonical ensemble MD
   simulation in dependence of $\invsupfraction - 1$,
   i.e.\ the ratio between air and \chem{CO_2}
   molecules in the system, for $\temperature$ $=$ $-44.8$ $^\circ$C
   with $\supersatrho$ $=$ $3.42$ and $\temperature$ $=$ $-34.8$ $^\circ$C
   with $\supersatrho$ $=$ $2.8$ as well as $\temperature$ $=$ $-23$ $^\circ$C
   with $\supersatrho$ $=$ $2.34$. Note that in the latter two cases, no
   nucleation was detected during the MD simulation run for $\invsupfraction - 1$
   = $1$ and $2$, implying that $\formationrate\slash\volume$ $<$ $10^{25}$ cm$^{-3}$s$^{-1}$.
}
\label{figpef}
\end{figure}

While this qualitatively
confirms CNT with the pressure effect, which leads to $\Weffect$ factors
on the order of $0.1$ for $\supfraction = 1\slash{}2$ and $0.01$ for
$\supfraction = 1\slash{}3$, the actual decrease of $\formationrate$
approaches the range $0.3$ to $0.4$ in the limit of large threshold sizes
for both values of $\supfraction$.
This impression consolidates itself if the results for $\temperature$ =
$-34.8$ and $-23$ $^\circ$C are also regarded, cf.\ Fig.\ \ref{figpef}
and Tab.\ \ref{tabnuc2}. The normalized \cite{WHBR08} pressure effect
$\Weffect$, corresponding to the deviation between $\CNT\nuclrate(\supfraction)$
and $\CNT\nuclrate(\supfraction = 1)$, decreases even faster with $\supfraction \to 0$ at high temperatures.
Qualitatively, this is confirmed by simulation results, e.g.\ no
nucleation at all was detected at these temperatures for $\supfraction = 1\slash{}3$,
which implies $\nuclrate < 10^{25}$ cm$^{-3}$s$^{-1}$. However, from the
available results for $\supfraction = 1\slash{}2$ it is evident that
the pressure effect is overestimated by CNT, in particular at
$\temperature = -23$ $^\circ$C.

%% Figure 3(h): J dependence on Y, comparison with Wedekind et al.
%

%% Table 3b(b): all results for J_i at T = 238.35 K and 250.2 K
%
\begin{table}[b!]  
\centering
\begin{tabular}{cccc|cc|clcc}
$\temperature$ [$^\circ$C] & $\supersatrho$ & $\supfraction$
                           & $\threshold$
                           & $\formationrate\slash\volume$ [cm$^{-3}$s$^{-1}$]
                           & $\gaseffect$ & $\NcritCNT$ & $\Weffect$ 
                           & $\CNT\nuclrate$ [cm$^{-3}$s$^{-1}$]
                           & $\CNT{\nuclrate\slash\committor}$ [cm$^{-3}$s$^{-1}$] \\ \hline
$-34.8$ & $2.80$ & $1/2$ & $\phantom{0}50$
      & $1.5 \mult 10^{27}$ & $\phantom{<\,} 2.7\phantom{0}$
      & $\phantom{0}66$ & $0.03$ 
      & $2.9 \mult 10^{24}$ & $2.6 \mult 10^{25}$ \\
      &      & & $\phantom{0}85$
      & $1.6 \mult 10^{26}$ & $\phantom{<\,} 0.76$
      & & 
      & & $3.3 \mult 10^{24}$ \\
      &      & $1\phantom{/1}$ & $\phantom{0}50$
      & $5.6 \mult 10^{26}$ & $\phantom{<\,} 1\phantom{.00}$
      & $\phantom{0}41$ &  $1$
      & $7.6 \mult 10^{25}$ & $9.9 \mult 10^{25}$ \\
      &      &   & $\phantom{0}85$        
      & $2.1 \mult 10^{26}$ & $\phantom{<\,} 1\phantom{.00}$
      & & 
      & & $7.6 \mult 10^{25}$ \\
      & $3.08$ & $1/2$ & $\phantom{0}50$
      & $5.5 \mult 10^{27}$ & $\phantom{<\,} 9.5\phantom{0}$
      & $\phantom{0}65$ & $0.02$
      & $3.9 \mult 10^{24}$ & $3.1 \mult 10^{25}$ \\
      &      & & $150$         
      & $3.1 \mult 10^{26}$ & $\phantom{<\,} 1.0\phantom{0}$
      & & 
      & & $3.9 \mult 10^{24}$ \\
      &      & $1\phantom{/1}$ & $\phantom{0}50$
      & $6.3 \mult 10^{27}$ & $\phantom{<\,} 1\phantom{.00}$
      & $\phantom{0}39$ & $1$
      & $1.3 \mult 10^{26}$ & $1.6 \mult 10^{26}$ \\
      &      & & $150$          
      & $2.9 \mult 10^{26}$ & $\phantom{<\,} 1\phantom{.00}$
      & & 
      & & $1.3 \mult 10^{26}$ \\
      & $3.36$ & $1/3$ & $\mathrm{n/a}$
      & $\phantom{0.0} < 10^{25}$ & ${<\,} 0.01$
      & $127$ & $4.2\mult10^{-6}$
      & $1.1 \mult 10^{21}$ & $\mathrm{n/a}$ \\
      &      & $1/2$ & $\phantom{0}50$
      & $1.1 \mult 10^{28}$ & $\phantom{<\,} 1.6\phantom{0}$
      & $\phantom{0}65$ & $0.02$
      & $4.2 \mult 10^{24}$ & $8.7 \mult 10^{24}$ \\
      &      & & $300$          
      & $3.2 \mult 10^{26}$ & $\phantom{<\,} 0.22$
      & & 
      & & $4.2 \mult 10^{24}$ \\
      &      & $1\phantom{/1}$ & $\phantom{0}50$
      & $6.7 \mult 10^{27}$ & $\phantom{<\,} 1\phantom{.00}$
      & $\phantom{0}37$ & $1$
      & $1.8 \mult 10^{26}$ & $2.1 \mult 10^{26}$ \\
      &      & & $300$         
      & $1.4 \mult 10^{27}$ & $\phantom{<\,} 1\phantom{.00}$
      & & 
      & & $1.8 \mult 10^{26}$ \\ \hline
$-23.0$ & $2.34$ & $1/2$ & $\phantom{0}50$
      & $1.1 \mult 10^{28}$ & $\phantom{<\,} 8.5\phantom{0}$
      & $140$ & $1.9\mult10^{-4}$
      & $4.0 \mult 10^{22}$ & $1.8 \mult 10^{27}$ \\
      &      & & $100$          
      & $1.1 \mult 10^{27}$ & $\phantom{<\,} 3.3\phantom{0}$ 
      & & 
      & & $7.8 \mult 10^{23}$ \\
      &      & $1\phantom{/1}$ & $\phantom{0}50$
      & $1.3 \mult 10^{27}$ & $\phantom{<\,} 1\phantom{.00}$
      & $\phantom{0}54$ & $1$
      & $1.4 \mult 10^{26}$ & $3.9 \mult 10^{26}$ \\
      &      & & $100$          
      & $3.4 \mult 10^{26}$ & $\phantom{<\,} 1\phantom{.00}$ 
      & & 
      & & $1.4 \mult 10^{26}$ \\
      & $2.53$ & $1/2$ & $\phantom{0}85$
      & $7.4 \mult 10^{27}$ & $\phantom{<\,} 3.4\phantom{0}$
      & $143$ & $1.0 \mult 10^{-4}$
      & $3.0 \mult 10^{22}$ & $3.9 \mult 10^{24}$ \\
      &      & & $200$          
      & $7.4 \mult 10^{26}$ & $\phantom{<\,} 0.96$
      & & 
      & & $3.1 \mult 10^{22}$ \\
      &      & $1\phantom{/1}$ & $\phantom{0}85$
      & $2.2 \mult 10^{27}$ & $\phantom{<\,} 1\phantom{.00}$
      & $\phantom{0}52$ & $1$
      & $1.9 \mult 10^{26}$ & $1.9 \mult 10^{26}$ \\
      &      & & $200$          
      & $7.7 \mult 10^{26}$ & $\phantom{<\,} 1\phantom{.00}$
      & & 
      & & $1.9 \mult 10^{26}$ \\
      & $2.72$ & $1/3$ & $\mathrm{n/a}$
      & $\phantom{0.0} < 10^{25}$ & ${<\,} 0.01$ 
      & $879$ & $4.3 \mult 10^{-25}$
      & $2.3 \mult 10^{2\phantom{0}}$ & $\mathrm{n/a}$ \\
      &        & $1/2$ & $\phantom{0}75$
      & $1.3 \mult 10^{28}$ & $\phantom{<\,} 2.6\phantom{0}$
      & $150$ & $4.2\mult10^{-5}$
      & $1.7 \mult 10^{22}$ & $1.8 \mult 10^{25}$ \\
      &      & & $250$          
      & $1.6 \mult 10^{27}$ & $\phantom{<\,} 1.1\phantom{0}$
      & & 
      & & $1.7 \mult 10^{22}$ \\
      &      & $1\phantom{/1}$ & $\phantom{0}75$
      & $4.8 \mult 10^{27}$ & $\phantom{<\,} 1\phantom{.00}$
      & $\phantom{0}50$ & $1$
      & $2.5 \mult 10^{26}$ & $2.6 \mult 10^{26}$ \\
      &      & & $250$          
      & $1.4 \mult 10^{27}$ & $\phantom{<\,} 1\phantom{.00}$ 
      & & 
      & & $2.5 \mult 10^{26}$ \\ \hline
\end{tabular}
\caption{
   \label{tabnuc2}
   \Droplet{} formation rate and carrier gas effect from YM canonical
   ensemble MD simulation as well as critical \droplet{} size (in
   molecules), normalized \cite{WHBR08} pressure effect, nucleation
   rate and \droplet{} formation rate according to the PE variant of CNT,
   in dependence of temperature, supersaturation and mole fraction
   of $\chem{CO_2}$ in the \vapour{} as well as the YM threshold size (in
   molecules).
}
\end{table}

%%%%%%%%%%%%%%%%%%%%%%%%%%%%%%%%%%%%%%%%%%%%%%%%
% hier beginnt der Teil zur Tropfenverdampfung %
%%%%%%%%%%%%%%%%%%%%%%%%%%%%%%%%%%%%%%%%%%%%%%%%

%T% \citep{HL08}  % weitere hier zu zitierende Simulationsstudie
%T% \citep{MKEY07}  % Ansatz von MKEY ist nur OK, wenn die richtige Fluktuation verwendet wird
%T% 
%T% %% Figure 4(h): vapor pressure
%T% %
%T% \begin{figure}[h!]
%T% \centering
%T% % \includegraphics[width=8.4cm]{fig/psn}
%T% \caption{
%T% }
%T% \label{figpsn}
%T% \end{figure}
%T% 
%T% %% Table 4(t): all results on the vapor pressure of droplets, including temperature effects
%T% %
%T% \begin{table}[t!]
%T% \centering
%T% \caption{
%T%    \label{tabdro}
%T%    \dots
%T% }
%T% \end{table}

\section{Conclusion}
\label{sec:cls}

%T% 
%T% \citep{HL08}  % Vergleichsaussage Tropfenverdampfung
%T% 

\Vapour-liquid coexistence in equilibrium and non-equilibrium was studied
by molecular simulation for systems consisting of \chem{CO_2}, \chem{N_2}, \chem{O_2},
and \chem{Ar}. For the nucleation of pure \chem{CO_2}, it was found that 
both the SC and the PE variant of CNT underestimate the nucleation rate by
up to a factor $20$ for SC and between one and three orders of magnitude
for PE.

It should be noted that this result for pure \chem{CO_2} nucleation
is both qualitatively and quantitatively similar to the
deviation of CNT for homogeneous nucleation of the truncated-shifted LJ
fluid \citep{HVH08}. In that case, it was shown that the deviation can
be corrected by a thermodynamically consistent approach that takes
both the curvature dependence of $\surfacetension$ and the increased
surface area due to fluctuations of the \droplet{} geometry into account.
Future work could elaborate on that and validate whether a theory based
on these considerations, e.g.\ as expressed by Eqs.\ (\ref{eqn:corr1})
and (\ref{eqn:corr2}), applies for \chem{CO_2} as well.

The effect of air as a carrier gas on \chem{CO_2} nucleation as
determined from MD simulation qualitatively confirms the theory
outlined by \cite{WHBR08}. Although significant quantitative deviations were found,
these are partly due to the fact that at the high densities corresponding
to the present simulations, air cannot be reliably described by
the ideal gas equation. The non-ideality leads to a lower total pressure
and thereby reduces the magnitude of the pressure effect to
a certain extent.

%%%%%%%%%%%%%%%%%%
% Acknowledgment %
%%%%%%%%%%%%%%%%%%

\section*{Acknowledgements}

The authors would like to thank
M.\ Bernreuther (Stuttgart) for his valuable support regarding organisational and technical issues,
D.\ Reguera L\'opez and J.\ Wedekind (Barcelona), G.\ Chkonia and J.\ W\"olk (Cologne),
G.\ Guevara Carri\'on and J.\ Walter (Kaiserslautern)
as well as Y.-L.\ Huang and S.\ Miroshnichenko (Paderborn)  % \footnote{\twn{黃佑霖} (Huang Yow-Lin)
for raising interest in the subject and for lively discussions,
and the Federal Ministry of Education and Research (BMBF) for funding the project IMEMO.
The presented research
was carried out under the auspices of the Boltzmann-Zuse Society of Computational
Molecular Engineering (BZS), and the simulations were performed on the
Opteron cluster \textit{phoenix} and the NEC Nehalem cluster \textit{laki}
at the High Performance Computing Center Stuttgart (HLRS) under the grant MMHBF.

\section*{Literature}
\label{sec:lit}

% \bibliographystyle{elsarticle-harv}
% \bibliography{co2air}

\end{document}